\documentclass[preprint,showpacs,preprintnumbers,amsmath,amssymb]{revtex4}
\usepackage{graphicx}

\begin{document}

\title{F\"orster mechanism of electron-driven proton pump}

\author{ Anatoly  Yu. Smirnov$^{1,2}$, Lev G. Mourokh$^{1,3,4}$, and Franco Nori$^{1,5}$}

\affiliation{ $^1$ Frontier Research System, The Institute of Physical and
Chemical Research (RIKEN), Wako-shi, Saitama, 351-0198, Japan \\
$^2$ CREST, Japan Science and Technology Agency, Kawaguchi,
Saitama, 332-0012, Japan \\
$^3$ Department of Physics, Queens College, The City University of
New York, Flushing, New York 11367, USA \\
$^4$ Department of Engineering Science and Physics, College of Staten Island, The City University of
New York, Staten Island, New York 10314, USA \\
$^5$ Center for Theoretical Physics, Physics Department, The University of Michigan, Ann Arbor, MI 48109-1040, USA}

\date{\today}

\begin{abstract}
{We examine a simple model of proton pumping through the inner membrane of mitochondria in the living cell. We demonstrate that
the pumping process can be described using approaches of condensed matter physics. In the framework of this model, we show that
the resonant F\"orster-type energy exchange due to electron-proton Coulomb interaction can provide an unidirectional flow of
protons against an electrochemical proton gradient, thereby accomplishing proton pumping. The dependence of this effect on
temperature as well as electron and proton voltage build-ups are obtained taking into account electrostatic forces and noise in
the environment. We find that the proton pump works with maximum efficiency in the range of temperatures and transmembrane
electrochemical potentials which correspond to the parameters of living cells.
 }
\end{abstract}


 \maketitle

\section{Introduction}

A living cell can be considered as a tiny electrical battery with a transmembrane potential difference of order $-70$ mV  (with a
negatively charged interior). Even a higher potential, $\Delta V \sim -200$ mV, is applied to the inner membrane of a
mitochondrion, an organelle, which produces most of the energy consumed by the cell. \cite{Alberts02,Wik04,Brand06}.
 To create and maintain such an electrical potential, mitochondria employ numerous proton pumps converting energy
of electrons into an electrochemical proton gradient that is harnessed thereafter to drive the synthesis of adenosine triphosphate
(ATP) molecules. Translocation of protons across the inner membrane of mitochondria is performed by the enzyme cytochrome $c$
oxidase (COX). Although crystal structure of COX is known in detail, a molecular mechanism of the redox-driven proton pumping
remains a mystery despite of the significant latest advances  based on time-resolved optical and electrometric measurements
\cite{Belev07,WV07}.

The electron transport chain of COX consists of four metal redox centers, ${\rm Cu}_A$, heme $a$, heme $a_3$, and ${\rm Cu}_B$
\cite{Brand06,Papa04,Bloch04}. The process starts when the mobile electron carrier, cytochrome $c$, moving from the positively
charged P-side of the membrane, donates a high-energy electron to a dinuclear copper site, ${\rm Cu}_A$ (see Fig.1). After that,
the electron proceeds to the heme $a$ with a subsequent transfer to the binuclear center formed  by heme $a_3$ and a copper ion
${\rm Cu}_B$, where the dioxygen molecule ${\rm O}_2$ is reduced to water. To produce two molecules of water in the catalytic
cycle with four electrons $(e^-)$ \cite{Stuch06}, $$ {\rm O}_2 + 8~ {\rm H}^+_{\rm N} + 4~ e^- \rightarrow 2~{\rm H}_2{\rm O} + 4~
{\rm H}^+_{\rm P},$$ the cytochrome oxidase consumes 4 substrate (chemical) protons which are translocated from the negative
N-side of the inner mitochondrion membrane to the binuclear center. In the process, four more protons (${\rm H}^+_{\rm N}$) are
taken from the N-side and pumped to the positive side (${\rm H}^+_{\rm P}$). Here, subscripts N and P for the protons denote the
location of the proton ${\rm H}^+$ at the negative (N) or positive (P) side of the membrane, respectively. A residue $E278$ (for
the $Paracoccus~ denitrificans$ enzyme) or a conserved glutamic acid, $Glu242$ (for the bovine enzyme \cite{WV07,WV06}), located
at the end of the so-called D-pathway \cite{BelNat06}, can serve as starting points for both substrate and pumped protons on their
way from the N-side to the binuclear center. In the next phase, a proton is transferred to an unknown yet protonable pump site $X$
which is located on the P-side of the heme groups and electrostatically coupled to heme $a$  and to the binuclear iron-copper
center $a_3/{\rm Cu}_{\rm B}$ \cite{Belev07,WV07}. On the final stage, the proton moves from the site X to the positive side of
the membrane after uphill pumping. In the context of a pure electrostatic model proposed in Refs. \cite{Belev07,WV07}, the
protonation of the site X leads to the equalization of electron energy levels in hemes $a$ and $a_3$ that facilitates a transfer
of an electron from heme $a$ to the binuclear center. This electron attracts a substrate proton which moves from the N-side of the
membrane to the site X, expelling the first, pre-pumped proton to the P-side. Detailed density functional and electrostatics
studies of this and other models have been performed in \cite{SiegJPC03,Stuch06,Blom06,Sieg07,Ols07,Hos06}. However, a mechanism
of {\it energy transmission from electrons to protons} resulting in an unidirectional translocation of protons against the
concentration gradient is still uncertain. For better understanding of this phenomenon, it is useful to combine a comprehensive
analysis of the energetic and spatial structure of enzymes with simple and physically transparent models.

In the present paper, we approach the problem taking into account the  similarity of the electron-driven proton transfer to the
quantum transport of electrons through nanostructures \cite{Wingr}. The interaction between electrons and protons is described by
a Coulomb potential, but, in addition to the standard electrostatic terms, we analyze effects of the F\"orster-type Coulomb
exchange \cite{Forst65}  on the resonant energy transduction between electron and proton subsystems. Each of the subsystems is
supposed to have two active sites: $1_e,2_e$ for electrons, and $1_p,2_p$ for protons. We consider here the possibility when both
electron sites belong to the same potential well, localized in the binuclear center $a_3/{\rm Cu}_{\rm B}$, while both active
proton states $2_p$ and $1_p$ can be ascribed to the pump center X (see Fig.1). This positioning of active sites corresponds in
some sense to the electrostatic model of Ref.~\cite{WV07}, based on time-resolved measurements of electron transfer in  COX enzyme
\cite{Belev07}.

During the F\"orster process, an electron moves from the state
$2_e$, which has a higher energy, to the state $1_e$, with a lower
energy; whereas a proton jumps from the lower-energy state $1_p$
to the higher-energy state $2_p$ (see Fig.~1). The same mechanism
is responsible for the Fluorescence Resonant Energy Transfer
(FRET) in biological systems \cite{ForstBio}, as well as for the
exciton transfer in condensed matter \cite{Klim04}.

The F\"orster term originates from the matrix element of the
Coulomb electron-proton potential between the overlapping  wave
functions of the electron states $2_e$ and $1_e$, and the
overlapping wave functions of proton states $1_p$ and $2_p$
\cite{Gov05}. Calculations show that this term is directly
proportional to the product of the dipole moments of electron and
proton two-level systems, also inversely proportional to the cube
of the distance between the electron and proton sites, and
requires to satisfy resonant conditions for the energies of the
electron and proton subsystems. Accordingly, the F\"orster term is
much weaker than standard electrostatic terms. However, as a
consequence of its overlapping origin, this term opens a new
channel for {\it simultaneous } tunneling of electrons and
protons, in addition to the {\it direct } tunneling. We
demonstrate that it is the {\it F\"orster-type coupling} that
results in an effective electron-proton energy transfer, followed
by the proton pumping from the negative to the positive side of
the inner mitochondria membrane.

The rest of the paper is structured as follows. Formulation of Hamiltonians and energetic spectra of the problem is presented in
Section II. Expressions for electron and proton currents are obtained in Section III. In Section IV, we derive equations of motion
for the density matrix. In Section V, these equations are solved numerically and the obtained dependencies of the proton current
on temperature, electron and proton voltage build-ups, and deviation from the resonant conditions are discussed. Section VI
contains our conclusions.

\section{Model Formulation}

Electrons and protons on sites $\sigma = 1,2 $ are characterized
by the Fermi operators $a_{\sigma}^+, a_{\sigma}$, and $
b_{\sigma}^+, b_{\sigma}$, respectively, with the corresponding
populations, $n_{\sigma} = a_{\sigma}^+a_{\sigma}$ and $N_{\sigma}
= b_{\sigma}^+b_{\sigma}$ (we interchangeably use the notation
``site" = ``state"). We assume that each electron site or proton
site can be occupied by a single particle, so the maximal
populations can be, at most, one electron on each one of the two
separate electron sites, and, at most, one proton on each one of
the two separate proton sites. To describe the continuous flow of
carriers through the system, we assume that the electron site 2 is
coupled to the left (L) reservoir, which serves as a source of
electrons, and the electron site 1 is coupled to the right
reservoir (R) playing the role of drain. At the same time, the
proton site 1 can be populated when protons jump from the
reservoir located on the negative (N) side of the membrane. On the
positive side of the membrane, there is another proton reservoir
which serves to depopulate of the proton site 2 (see Fig.~1b). In
the framework of this model, here we neglect the couplings between
the electron site 1 and the reservoir L, and between the site 2
and the reservoir R. We also neglect the tunneling between the
proton site 1 and the positive side of the membrane (P), as well
as the tunneling between the proton site 2 and the negative side
of the membrane (N).

The electrons in the reservoir (lead) $\alpha$ ($\alpha = L,R$) or
the protons in the reservoir (lead) $\beta$ ($\beta = N,P$) can be
characterized by additional parameters $k$ and $q$, respectively,
which have meanings of wave vectors in condensed matter physics.
To describe the electronic and protonic sources and drains, we
introduce the electron creation and annihilation operators in the
$\alpha$-lead as $c_{k\alpha}^+, c_{k\alpha}$, and their proton
counterparts for the $\beta$-lead as $d_{q\beta}^+, d_{q\beta}$.
The number of electrons in the $\alpha$-lead is determined by the
operator $ \sum_k n_{k\alpha}$, with $n_{k\alpha}
=c_{k\alpha}^+c_{k\alpha},$ whereas the proton population of the
$\beta$-lead is given by the operator $ \sum_q N_{q\beta}$, with
$N_{q\beta}= d_{q\beta}^+d_{q\beta}$. It is well-known that in
real biological structures, couplings between the active sites
$1,2$ and the reservoirs can be mediated by many bridge states,
similar to the ${\rm Cu}_{\rm A}$-site and heme $a$, which can be
subjected to conformational changes. Conformation changes can also
provide a selectivity in coupling between the active sites and the
leads \cite{Wik04}.

\subsection{Electron and proton Hamiltonians}

The Hamiltonian of the electron-proton system incorporates a term related to eigenenergies $\epsilon_{\sigma}^{(0)},
E_{\sigma}^{(0)}$ of electrons and protons, respectively, located on the sites $\sigma = 1,2$,  as well as  a term describing
electron and proton energies $\epsilon_{k\alpha}, E_{k\beta}$ of the leads $\alpha = L,R; \beta = N,P:$
\begin{equation}
H_{{\rm init}} = \sum_{\sigma} ( \epsilon_{\sigma}^{(0)} n_{\sigma} + E_{\sigma}^{(0)} N_{\sigma} ) + \sum_{k\alpha}
\epsilon_{k\alpha} c_{k\alpha}^+ c_{k\alpha} + \sum_{q\beta} E_{q\beta} d_{q\beta}^+d_{q\beta}. \label{Hin}
\end{equation}
The Hamiltonian $H_{\rm dir}$,
\begin{equation}
H_{\rm dir} = - \Delta_a a_2^+a_1 - \Delta_a^* a_1^+a_2 - \Delta_b b_2^+b_1 - \Delta_b^* b_1^+b_2, \label{Hdir}
\end{equation}
is responsible for the direct tunneling of electrons and protons between the corresponding sites 1 and 2, with the rates
$\Delta_a$ and $\Delta_b.$ Notice that the direct tunneling has a highly non-resonant character since the energy levels of the
sites 1 and 2 are well separated: $\epsilon_{2}^{(0)}-\epsilon_{1}^{(0)} \gg \Delta_a, ~E_2^{(0)} - E_1^{0)} \gg \Delta_b.$ To
take into consideration the coupling of the active sites 1 and 2 to the corresponding reservoirs of electrons and protons, we
introduce the tunneling Hamiltonian
\begin{equation}
H_{\rm tun} = - \sum_k t_{kR} c_{kR}^+a_1 -\sum_k t_{kL} c_{kR}^+a_2 - \sum_q T_{qN} d_{qN}^+b_1 -  \sum_q T_{qP} d_{qP}^+b_2 +
h.c. \label{Htun}
\end{equation}

The Coulomb force plays the most important role in the process of energy transfer from the electron subsystem to protons. This
interaction is determined by the Coulomb potential
\begin{equation}
u(r_e,r_p, R ) = - \; \frac{e^2}{4\pi \epsilon_0 \epsilon_r | r_p - r_e + R |}, \label{UPot}
\end{equation}
where $r_e,r_p$ are the electron and proton positions in their
local frame of reference, and $R$ is the distance between the
electron and proton sites, $ R \gg r_e , r_p$. A direct
electron-proton Coulomb attraction is determined by the energies
$u_{\sigma \sigma'}$ $(\sigma = 1_e,2_e; \ \sigma'=1_p,2_p).$ In
addition, we take into account the repulsion of the two electrons
located at the sites $1_e$ and $2_e$ (energy scale $\sim u_{e}$)
jointly with the repulsion of two protons localized on the sites
$1_p$ and $2_p$ (an energy parameter $u_p$). It should be noted
that all energy characteristics $u_{\sigma\sigma'},u_e,u_p$ are
modified compared to their original values because of Coulomb
interactions between the active sites and the electron and proton
reservoirs. As a result, the Hamiltonian related to the direct
Coulomb interaction has the form
\begin{equation}
H_{C}^{(0)} = - \sum_{\sigma\sigma'} u_{\sigma\sigma'} n_{\sigma}N_{\sigma'} + u_e n_1 n_2 + u_p N_1 N_2. \label{HC0}
\end{equation}

\subsection{F\"orster term }

The direct Coulomb coupling between electrons and protons should be complemented by the F\"orster term,
\begin{equation}
H_F = V_F a_1^+a_2 b_2^+b_1 + V_F^* a_2^+a_1 b_1^+b_2 , \label{HF}
\end{equation}
 which originates from the cross matrix element of the Coulomb potential (\ref{UPot})
\begin{equation}
V_F = - \langle 1_e 2_p |\; \frac{e^2}{ 4\pi \epsilon_0 \epsilon_r |r_p -r_e + R| }\; | 2_e 1_p \rangle. \label{VF1}
\end{equation}
This matrix element is taken over the electron-proton wave function $|1_e 2_p \rangle$, with the electron being in the state $1_e$
and the proton being in the state $2_p$, and the wave function $ | 2_e 1_p \rangle $, with the electron being in the state $2_e$
and the proton being in the state $1_p$. The F\"orster term can be significant in the case of an electron-proton resonance when
the distance between the electron energy levels $\epsilon_1$ and $\epsilon_2$ is close to the separation of the proton energy
levels $E_1$ and $E_2: \epsilon_2 - \epsilon_1 \simeq E_2 - E_1.$ Therefore, the states $|1_e 2_p \rangle$ and $ | 2_e 1_p \rangle
$ have almost the same energy:  $\epsilon_1 + E_2 \simeq \epsilon_2 + E_1,$ that is favorable
 to transitions between these states.
 The contributions  of the other cross-elements of the electron-proton Coulomb
 attraction, such as
 $  \langle 2_e 1_p | u(r_e,r_p,R)  | 1_e 2_p \rangle,
 \langle 2_e 2_p | u(r_e,r_p,R)  | 1_e 2_p \rangle, $ etc.,  which have
a non-resonant character, are quite small \ ($\sim V_F/ (E_2 - E_1) \ll 1$ at $E_2-E_1 \sim 500 $ meV, $V_F\sim 1 $ meV), and can
be neglected. We consider here a situation where the wave functions $1_e,2_e$ represent the ground and the first excited state of
the electron in a parabolic potential well which is placed a distance $R$ from the proton potential well containing two proton
states $1_p,2_p$. Using the expansion ($ r = |\textbf{r}| \ll R= |\textbf{R}| $),
\begin{equation}
\frac{1}{|\textbf{R} - \textbf{r}|} = \frac{1}{R}\left[ 1 - \frac{\textbf{r}\cdot\textbf{R}}{R^2} + 3
\frac{(\textbf{r}\cdot\textbf{R})^2}{R^4}  - \frac{r^2}{R^2} + ...\right], \label{Expan}
\end{equation}
 we find that the matrix element $V_F$ characterizing the
strength of the F\"orster term is proportional to the product of the dipole moments, $er_0$ and $eR_0$, of the electron and proton
sites 1 and 2 and inversely proportional to the cubic power of the distance $R$ between these sites:
\begin{equation}
V_F = \frac{e^2}{2\pi \epsilon_0 \epsilon_r} \frac{r_0 R_0}{R^3}. \label{VF2}
\end{equation}
For a protein with a dielectric constant $\epsilon_r = 3$ and the electron/proton wave function spreadings $r_0 =0.1$ nm and
$R_0=0.01$ nm, we estimate the F\"orster matrix element as $V_F\simeq 1$ meV, if the distance between the electron and proton
sites $R=1$ nm.

\subsection{Dissipative environment}

To account for the effects of a dissipative environment on the electron and proton transfer, we resort to the well-known model
\cite{Garg85,Krish01,MarSut85} where the polar medium surrounding the electron and proton active sites is represented by two
systems of harmonic oscillators with the following Hamiltonian:
\begin{eqnarray}
H_B = \sum_j \left( \frac{p_j^2}{2m_j} + \frac{m_j\omega_j^2 x_j^2}{2} \right)  +
\sum_j  \frac{m_j\omega_j^2 x_{j0} x_j}{2}  (n_2-n_1) + \nonumber\\
  \sum_j \left( \frac{P_j^2}{2M_j} + \frac{M_j\Omega_j^2 X_j^2}{2} \right)  +
  \sum_j  \frac{M_j\Omega_j^2 X_{j0} X_j}{2}  (N_1-N_2) . \label{HB1}
\end{eqnarray}
Here $\{x_j,p_j\}$ are positions and momenta of the oscillators coupled to the electron subsystem, whereas the variables $\{ X_j,
P_j\}$ are related to the proton environment. The electron and proton surroundings are characterized by their own sets of
effective masses $m_j$ and $M_j$ as well as by the two sets of eigenfrequencies $\omega_j $ and $ \Omega_j$. The strengths of the
couplings to the environments are determined by the shifts $x_{j0}$ and $X_{j0}$ of the equilibrium positions of the corresponding
$j$th-oscillator. The bath Hamiltonian, Eq.~(\ref{HB1}), can be rewritten in the form
\begin{eqnarray}
H_B = \sum_j \left( \frac{p_j^2}{2m_j} + \frac{m_j\omega_j^2 [ x_j +
(1/2)x_{j0}(n_2-n_1)]^2 }{2} \right)  + \nonumber\\
  \sum_j \left( \frac{P_j^2}{2M_j} + \frac{M_j\Omega_j^2 [ X_j +
  (1/2) X_{j0}(N_1-N_2)]^2 }{2} \right)  - \nonumber\\
\frac{1}{4} \lambda_a (n_1 + n_2) - \frac{1}{4} \lambda_b (N_1 + N_2) , \label{HB2}
\end{eqnarray}
where the parameters $\lambda_a$ and $\lambda_b $ are reorganization energies for the electron and proton environments,
\begin{equation}
\lambda_a =  \sum_j  \frac{m_j\omega_j^2 x_{j0}^2}{2}, \; \;  \lambda_b = \sum_j  \frac{M_j\Omega_j^2 X_{j0}^2}{2}.
\label{lambdaAB}
\end{equation}
The systems of independent harmonic oscillators are conveniently characterized by the spectral functions   $J_a(\omega)$ and
$J_b(\omega)$, defined as
\begin{equation}
J_a(\omega) = \sum_j  \frac{m_j\omega_j^3 x_{j0}^2}{2} \delta(\omega - \omega_j),    ~J_b(\omega) = \sum_j  \frac{m_j\Omega_j^3
X_{j0}^2}{2} \delta(\omega - \Omega_j), \label{Jab}
\end{equation}
so that
\begin{equation}
\lambda_{a} = \int_0^{\infty} \frac{d\omega}{\omega} J_{a}(\omega), \; \; \lambda_{b} = \int_0^{\infty} \frac{d\omega}{\omega}
J_{b}(\omega). \label{JLam}
\end{equation}

\subsection{Total Hamiltonian}

The total Hamiltonian of the system incorporates all the
above-mentioned terms, as
\begin{eqnarray}
H = H_0 + \sum_{k\alpha} \epsilon_{k\alpha} c_{k\alpha}^+c_{k\alpha} + \sum_{q\beta} E_{q\beta} d_{q\beta}^+d_{q \beta}
+ V_F a_1^+a_2b_2^+b_1 + V_F^* a_2^+a_1b_1^+b_2 - \nonumber\\
\Delta_a a_2^+a_1 - \Delta_a^* a_1^+a_2 - \Delta_b b_2^+b_1 -
\Delta_b^* b_1^+b_2 - \nonumber\\
 - \sum_k t_{kR} c_{kR}^+a_1 -
\sum_k t_{kR}^* a_1^+c_{kR} - \sum_k t_{kL} c_{kL}^+a_2
 -\sum_k t_{kL}^*  a_2^+ c_{kL} - \nonumber\\
\sum_q T_{qN} d_{qN}^+b_1 - \sum_q T_{qN}^*b_1^+ d_{qN} -   \sum_q T_{qP} d_{qP}^+b_2  -
 \sum_q T_{qP}^*b_2^+ d_{qP} + \nonumber\\
 \sum_j \left( \frac{p_j^2}{2m_j} + \frac{m_j\omega_j^2
 [ x_j + (1/2)x_{j0}(n_2-n_1)]^2 }{2} \right)  + \nonumber\\
  \sum_j \left( \frac{P_j^2}{2M_j} + \frac{M_j\Omega_j^2
 [ X_j + (1/2) X_{j0}(N_1-N_2)]^2 }{2} \right), \label{Ham1}
\end{eqnarray}
where the Hamiltonian
\begin{eqnarray}
H_0 = \sum_{\sigma} ( \epsilon_{\sigma} n_{\sigma} + E_{\sigma} N_{\sigma} ) - \sum_{\sigma\sigma'} u_{\sigma\sigma'}
n_{\sigma}N_{\sigma'} + u_e n_1 n_2 + u_p N_1 N_2 \label{H01}
\end{eqnarray}
is characterized by the renormalized energy levels,
$$\epsilon_{\sigma} = \epsilon_{\sigma}^{(0)} -(1/4)\lambda_a, \quad
 E_{\sigma} = E_{\sigma}^{(0)}  - (1/4) \lambda_b. $$
  Here the repulsion potentials, $u_e$ and $u_p$, also incorporate shifts
 proportional to the corresponding reorganization energies, $\lambda_a/2$ and $\lambda_b/2$.
 With the unitary transformation, $ \hat{U} = \hat{U}_a
\hat{U}_b$, where
$$\hat{U}_a = \exp[-(i/2)\sum_j p_jx_{j0}(n_1-n_2)], \quad \hat{U}_b =
\exp[-(i/2)\sum_j P_jX_{j0}(N_2-N_1)],$$ we can transform the
Hamiltonian $H$, Eq.~(\ref{Ham1}), to the form
\begin{eqnarray}
H = H_0 + \sum_{k\alpha} \epsilon_{k\alpha} c_{k\alpha}^+c_{k\alpha} + \sum_{q\beta} E_{q\beta} d_{q\beta}^+d_{q \beta} +
V_F a_1^+a_2b_2^+b_1 e^{i\xi} + V_F^* e^{-i \xi}  a_2^+a_1b_1^+b_2 - \nonumber\\
- \Delta_a e^{-i \xi_a} a_2^+a_1 - \Delta_a^* a_1^+a_2 e^{i \xi_a} - \Delta_b b_2^+b_1 e^{i \xi_b} - \Delta_b^* e^{-i \xi_b}
b_1^+b_2
- \nonumber\\
 \sum_k t_{kR}e^{-\frac{i}{2}\xi_a} c_{kR}^+a_1 -  \sum_k t_{kR}^*
a_1^+c_{kR}e^{\frac{i}{2}\xi_a} - \sum_k t_{kL} c_{kL}^+a_2 e^{\frac{i}{2}\xi_a}
 -\sum_k t_{kL}^* e^{-\frac{i}{2}\xi_a} a_2^+ c_{kL} - \nonumber\\
\sum_q T_{qN} d_{qN}^+b_1e^{\frac{i}{2}\xi_b} - \sum_q T_{qN}^*e^{-\frac{i}{2}\xi_b}b_1^+ d_{qN} -   \sum_q
T_{qP}e^{-\frac{i}{2}\xi_b} d_{qP}^+b_2
 -\sum_q T_{qP}^*b_2^+ d_{qP}e^{\frac{i}{2}\xi_b}  + \nonumber\\
 \sum_j \left( \frac{p_j^2}{2m_j} + \frac{m_j\omega_j^2
 x_j^2 }{2} \right)  +
  \sum_j \left( \frac{P_j^2}{2M_j} + \frac{M_j\Omega_j^2
  X_j^2 }{2} \right), \label{Ham2}
\end{eqnarray}
where  $$ \xi_a =
 (1/\hbar)\sum_j p_j \; x_{j0}, \quad \xi_b = (1/\hbar)\sum_j P_j \; X_{j0},$$
are stochastic phases operators, and  $ \xi = \xi_a + \xi_b.$ The
result of this transformation follows from the fact that, for an
arbitrary function $ \Phi[x_j,X_j] $, the operator $\hat{U}$
produces a shift of the oscillator's positions: $$ \hat{U}^+
\Phi[x_j,X_j] \hat{U} = \Phi[x_j + (1/2)x_{j0}(n_1-n_2), \; X_j +
(1/2)X_{j0}(N_2-N_1)].$$ In addition, this transformation results
in phase factors for electron and proton amplitudes:
$$ \hat{U}_a^+ a_1 \hat{U}_a =
 e^{-(i/2)\xi_a} a_1, \quad
 \hat{U}_a^+ a_2 \hat{U}_a =
 e^{(i/2)\xi_a} a_2, $$ and
 $$ \hat{U}^+ b_1 \hat{U} =
 e^{(i/2)\xi_b} b_1, \quad \hat{U}^+ b_2 \hat{U}_b =
 e^{-(i/2)\xi_b} b_2. $$

\subsection{Combined electron-proton eigenstates and energy eigenvalues}

The electron-proton system with no leads can be characterized by 16 basis states of the Hamiltonian $H_0$:
\begin{eqnarray}
 |1\rangle = |{\rm Vac}\rangle, ~|2\rangle = a_1^+|{\rm Vac}\rangle, ~|3\rangle = a_2^+|{\rm Vac}\rangle, ~|4\rangle = b_1^+|{\rm Vac}\rangle, ~|5\rangle =
b_2^+|{\rm Vac}\rangle, \nonumber\\
|6\rangle = a_1^+ b_1^+|{\rm Vac}\rangle, ~|7\rangle = a_1^+b_2^+|{\rm Vac}\rangle, ~|8\rangle = a_2^+b_1^+|{\rm Vac}\rangle,
~|9\rangle =
a_2^+b_2^+|{\rm Vac}\rangle, \nonumber\\
|10\rangle = a_1^+a_2^+|{\rm Vac}\rangle, ~|11\rangle = a_1^+a_2^+b_1^+|{\rm Vac}\rangle, ~|12\rangle = a_1^+a_2^+b_2^+|{\rm
Vac}\rangle, ~|13\rangle
= b_1^+b_2^+|{\rm Vac}\rangle, \nonumber\\
|14\rangle = a_1^+b_1^+b_2^+|{\rm Vac}\rangle,
 ~|15\rangle = a_2^+b_1^+b_2^+|{\rm Vac}\rangle, ~|16\rangle =
a_1^+a_2^+b_1^+b_2^+|{\rm Vac}\rangle. \label{basis1}
\end{eqnarray}
Here, $|{\rm Vac}\rangle$ represents the vacuum state, when both
electron active sites and both proton sites are empty, whereas,
for example, the state $|7\rangle = a_1^+b_2^+|{\rm Vac}\rangle $
corresponds to the case when one electron is located on the site
$1_e$ and one proton is located on the site $2_p$. The state $
|8\rangle = a_2^+b_1^+|{\rm Vac}\rangle $ is related to the
opposite situation with a single electron on the site $2_e$ and
one proton on the site $1_p$. It should be also noted that any
arbitrary operator ${\cal A}$ of the electron-proton system  can
be represented as an expansion in terms of the basis Heisenberg
matrices $\rho_m^n = |m\rangle \langle n| ~(m,n = 1,..,16)$:
${\cal A} = \sum_{m,n} {\cal A}_{mn} \rho_m^n.$ We will also use
notations $\rho_m \equiv \rho_m^m$ for the diagonal operator.
Thus, the operators $\{a_1,a_2,b_1,b_2\}$ can be represented as
\begin{eqnarray}
a_1 = \rho_1^2 + \rho_4^6 + \rho_5^7 + \rho_3^{10} + \rho_8^{11} + \rho_9^{12} + \rho_{13}^{14} + \rho_{15}^{16}, \nonumber\\
a_2 = \rho_1^3 + \rho_4^8 + \rho_5^9 - \rho_2^{10} - \rho_6^{11} - \rho_7^{12} + \rho_{13}^{15} - \rho_{14}^{16}, \nonumber\\
b_1 = \rho_1^4 + \rho_2^6 + \rho_3^{8} + \rho_{10}^{11} + \rho_5^{13} + \rho_7^{14}  + \rho_{9}^{15} + \rho_{12}^{16}, \nonumber\\
b_2 = \rho_1^5 + \rho_2^7 + \rho_3^{9} + \rho_{10}^{12} - \rho_4^{13} - \rho_6^{14}  - \rho_{8}^{15} - \rho_{11}^{16}.
\label{ab12}
\end{eqnarray}
The F\"orster operator in the Hamiltonian $H$, Eq.~(\ref{Ham2}), given by $ a_1^+a_2 b_2^+b_1$, is responsible for the electron
transition {\it from } the electron site $2_e$ {\it to } the site $1_e$ accompanied by the {\it simultaneous} proton transfer {\it
from } the proton site $1_p$ {\it to } the site $2_p$. In the basis introduced above, the F\"orster process corresponds to the
transition of the electron-proton system from the state $|8\rangle$ to the state $|7\rangle: ~ a_1^+a_2 \; b_2^+b_1 =
|7\rangle\langle 8| = \rho_7^8.$ Using the eigenfunctions, Eq.(\ref{basis1}), we can rewrite the Hamiltonian $H_0$ in a simple
diagonal form:
\begin{equation}
H_0 = \sum_{m=1}^{16} \varepsilon_m ~\rho_m, \label{H0Rho}
\end{equation}
 with the following energy spectrum:
\begin{eqnarray}
\varepsilon_1 =0, ~\varepsilon_2 =\epsilon_1, ~\varepsilon_3 = \epsilon_2, ~\varepsilon_4 = E_1, \nonumber\\
~\varepsilon_5 = E_2, ~\varepsilon_6 = \epsilon_1 + E_1 - u_{11}, \nonumber\\
\varepsilon_7 = \epsilon_1 + E_2 - u_{12}, ~\varepsilon_8 = \epsilon_2 + E_1 - u_{21}, \nonumber\\
~\varepsilon_9 = \epsilon_2 + E_2 - u_{22},
~\varepsilon_{10} = \epsilon_1 + \epsilon_2 + u_{e}, \nonumber\\
\varepsilon_{11} = \epsilon_1+\epsilon_2 + E_1 - u_{11}- u_{21} +u_e, \nonumber\\~\varepsilon_{12} = \epsilon_1+\epsilon_2 + E_2 -
u_{12}- u_{22}
+u_e, \nonumber\\
\varepsilon_{13} = E_1+ E_2 + u_p, ~\varepsilon_{14} =
\epsilon_1+E_1 + E_2 - u_{11}- u_{12} +u_p, \nonumber\\
 \varepsilon_{15} = \epsilon_2+E_1 + E_2 - u_{21}- u_{22} +u_p, \nonumber\\
 \varepsilon_{16} =
\epsilon_1+\epsilon_2+E_1 + E_2 -u_{11} -u_{12} - u_{21}- u_{22} + u_e + u_p. \label{spectr1}
\end{eqnarray}
For the F\"orster component of the Hamiltonian $H_F$, and for the
Hamiltonian $H_{\rm dir}$ describing the direct tunneling between
the sites $1_e,2_e$ and $1_p,2_p$, we obtain the expressions
\begin{equation}
H_F = V_F ~\rho_7^8 ~e^{i\xi} + V_F^*~ e^{-i\xi} ~\rho_8^7 \label{HF1}
\end{equation}
and
\begin{eqnarray}
H_{\rm dir} = -\Delta_a ~e^{-i \xi_a} ~( \rho_3^2 + \rho_8^6  + \rho_9^7 + \rho_{15}^{14} ) -
 \Delta_a^*  ( \rho_2^3 + \rho_6^8  + \rho_7^9 + \rho_{14}^{15} ) ~e^{i \xi_a} - \nonumber\\
 \Delta_b( \rho_5^4 + \rho_7^6  + \rho_9^8 + \rho_{12}^{11} ) ~e^{i \xi_b}
 -\Delta_b^* ~e^{-i \xi_b} ~(
\rho_4^5 + \rho_6^7  + \rho_8^9 + \rho_{11}^{12} ). \label{Hdir1}
\end{eqnarray}
It should be noted that the operators $H_F$ and $H_{\rm dir}$ are non-diagonal.

\section{Electron and proton currents}

The transfer of electrons (protons) can be quantitatively characterized by the particle current flows between left/right
(negative/positive) reservoirs, $i_{\alpha}$ ($I_{\beta}$), which are defined as
\begin{equation}
i_{\alpha} = \frac{d}{dt} \sum_k \langle c_{k\alpha}^+c_{k\alpha}\rangle, \quad I_{\beta} = \frac{d}{dt} \sum_q \langle
d_{q\beta}^+d_{q\beta}\rangle,  \label{CurDef}
\end{equation}
with indices $\alpha = L,R $ and $\beta = N,P.$ Taking into account the equations for electron and protons amplitudes in the
leads,
\begin{eqnarray}
i ~\dot{c}_{kL} = \epsilon_{kL} ~c_{kL} - t_{kL} ~a_2 ~e^{\frac{i}{2} \xi_a}, \nonumber\\ i ~\dot{c}_{kR} = \epsilon_{kR} ~c_{kL}
- t_{kR}~e^{-\frac{i}{2}
\xi_a} ~a_1, \nonumber\\
i ~\dot{d}_{qN} = E_{qN} ~d_{qN} - T_{qN}  ~b_1~ e^{\frac{i}{2} \xi_b}, \nonumber\\ i ~\dot{d}_{qP} = E_{qP} ~d_{qP} - T_{qP}
~e^{-\frac{i}{2} \xi_b}~ b_2, \label{cdEq}
\end{eqnarray}
we obtain for the currents,
\begin{eqnarray}
i_L = i \sum_k t_{kL} \langle c_{kL}^+ a_2 e^{\frac{i}{2} \xi_a} \rangle  + h.c.; ~i_R = i \sum_k t_{kR} \langle e^{-\frac{i}{2}
\xi_a} c_{kR}^+ a_1 \rangle + h.c. ; \nonumber\\
I_N = i \sum_q T_{qN} \langle d_{qN}^+ b_1 e^{\frac{i}{2} \xi_b} \rangle  + h.c.; ~I_P = i \sum_q T_{qP} \langle e^{-\frac{i}{2}
\xi_b} d_{qP}^+ b_2 \rangle + h.c. \label{Cur1}
\end{eqnarray}
It follows from Eq.~(\ref{cdEq}) that the leads' responses are described by the formulas
\begin{eqnarray}
c_{kL} = c_{kL}^{(0)} - t_{kL} \int dt_1 \; g_{kL}^r (t,t_1)\; a_2(t_1)\;
e^{\frac{i}{2} \xi_a(t_1)}, \nonumber\\
d_{qN} = d_{qN}^{(0)} - T_{qN} \int dt_1 \; g_{qN}^R (t,t_1)\; b_1(t_1) \;e^{\frac{i}{2} \xi_b(t_1)}, \label{cdReact}
\end{eqnarray}
etc., where $$g_{k\alpha}^r (t,t_1) = -i\; e^{-i\epsilon_{k\alpha}(t-t_1)} \; \theta(t-t_1), \quad g_{q\beta}^R (t,t_1) = -i
\;e^{-iE_{q\beta}(t-t_1)}\; \theta(t-t_1) $$ are the retarded Green functions of electrons and protons in the leads,
$c_{k\alpha}^{(0)},d_{q\beta}^{(0)}$ are unperturbed electron and proton operators in the electron reservoir $\alpha$ and  in the
proton lead $\beta$, respectively, and $ \theta(\tau)$ is the Heaviside step function. Within our model, we assume that
 electrons and protons in the leads are characterized by the Fermi distributions
$$ f_{\alpha}(\epsilon_{k\alpha}) = \left[
\exp\left(\frac{\epsilon_{k\alpha}- \mu_{\alpha}}{T} \right) + 1 \right]^{-1}, \quad F_{\beta}(E_{q\beta}) = \left[
\exp\left(\frac{E_{q\beta}- \mu_{\beta}}{T} \right) + 1 \right]^{-1},$$ respectively, having the same temperature $T ~(k_B=1)$.
However, the chemical potentials of electrons in the left $(\mu_L)$ and in the right $(\mu_R)$ lead, as well as chemical
potentials of the protons from the negative side of the membrane $(\mu_N)$ and from the positive one $(\mu_P)$, can be different
in the non-equilibrium case:
$$\mu_L = \mu_a +V_e, ~\mu_R = \mu_a, \quad \mu_N = \mu_b, ~\mu_P = \mu_b + V_p,$$
 where $V_e$ and
$V_p$ are electron and proton voltage build-ups, $\mu_a$ and $\mu_b$ are equilibrium chemical potentials of the electron and
proton reservoirs, respectively. Notice that the absolute value of the electron charge, $|e|$, is included into the definitions of
voltages $V_e,~V_p$, which are measured here in millielectronVolts (meV). Thus, the correlators of the unperturbed operators are
given by
\begin{eqnarray}
\langle c_{k\alpha}^{(0)+}(t) c_{k\alpha}^{(0)}(t_1)\rangle = f_{k\alpha}(\epsilon_{k\alpha}) \; e^{i \epsilon_{k\alpha}(t-t_1)},
\nonumber\\ \langle d_{q\beta}^{(0)+}(t) d_{q\beta}^{(0)}(t_1)\rangle = F_{q\beta}(E_{q\alpha}) \; e^{i E_{q\alpha}(t-t_1)}.
\label{cdCorr}
\end{eqnarray}
In the wide-band limit, it is convenient to introduce frequency-independent densities of electron (proton) states,
$\gamma_{\alpha} ~( \Gamma_{\beta} )$, as
\begin{equation}
\gamma_{\alpha} = 2\pi \sum_k |t_{k\alpha}|^2 \delta (\omega - \epsilon_{k\alpha}); ~ \Gamma_{\beta} = 2\pi \sum_q |T_{q\beta}|^2
\delta (\omega - E_{q\beta}). \label{gamLeads}
\end{equation}
It should be noted that the currents $i_{\alpha}$ and $I_{\beta}$ are involved in the equations for the averaged populations
derived from the Hamiltonian, Eq.~(\ref{Ham2}),
\begin{eqnarray}
 \langle \dot{n}_1\rangle = - i  V_F \langle a_1^+a_2 b_2^+b_1
e^{i\xi}\rangle  + i V_F^* \langle e^{-i\xi}a_2^+a_1 b_1^+b_2 \rangle +
i \Delta_a^* \langle a_1^+a_2 e^{i\xi_a}\rangle - i\Delta_a \langle e^{-i\xi_a} a_2^+a_1\rangle - i_R; \nonumber\\
\langle \dot{n}_2 \rangle =  i V_F \langle a_1^+a_2 b_2^+b_1 e^{i\xi}\rangle  - i V_F^* \langle e^{-i\xi}a_2^+a_1 b_1^+b_2\rangle
+ i\Delta_a \langle e^{-i\xi_a} a_2^+a_1\rangle   - i\Delta_a^* \langle a_1^+a_2 e^{i\xi_a}\rangle
- i_L; \nonumber\\
 \langle \dot{N}_1 \rangle  =  i V_F \langle a_1^+a_2 b_2^+b_1
e^{i\xi}\rangle - i V_F^* \langle e^{-i\xi}a_2^+a_1 b_1^+b_2\rangle + i \Delta_b^* \langle e^{-i\xi_b} b_1^+b_2\rangle  -
i\Delta_b \langle b_2^+b_1 e^{i\xi_b}\rangle  -
I_N; \nonumber\\
 \langle \dot{N}_2 \rangle =  -i V_F \langle a_1^+a_2 b_2^+b_1
e^{i\xi}\rangle + i V_F^* \langle e^{-i\xi}a_2^+a_1 b_1^+b_2\rangle + i \Delta_b \langle b_2^+b_1 e^{i\xi_b}\rangle   -
i\Delta_b^* \langle e^{-i\xi_b} b_1^+b_2\rangle - I_P. \label{nN12}
\end{eqnarray}
Here, the brackets $\langle ..\rangle $ denote averaging over the equilibrium states of electron and proton reservoirs,
complemented by the averaging over fluctuations of both dissipative environments. It is evident that in the steady-state regime,
when the time derivatives of all populations are zero, the electron and proton currents
 are determined by the F\"orster process and by the direct tunneling:
 \begin{eqnarray}
 i_L = - i_R = i V_F \langle a_1^+a_2b_2^+b_1 \, e^{i\xi}\rangle  -
 i V_F^* \langle e^{-i\xi}\,a_2^+a_1b_1^+b_2 \rangle + \nonumber\\
 i \Delta_a \langle
 e^{-i\xi_a} \, a_2^+a_1\rangle - i\Delta_a^*\langle a_1^+a_2
 \, e^{i\xi_a}\rangle, \nonumber\\
 I_N = - I_P = i V_F \langle a_1^+a_2b_2^+b_1 \, e^{i\xi}\rangle  -
 i V_F^* \langle e^{-i\xi}\, a_2^+a_1b_1^+b_2 \rangle + \nonumber\\
 i\Delta_b^*\langle
 e^{-i\xi_b} \, b_1^+b_2\rangle - i\Delta_b \langle b_2^+b_1
 \, e^{i\xi_b}\rangle. \label{Cur3}
\end{eqnarray}
We assume that the F\"orster energy $V_F$, the direct tunneling rates, $\Delta_a$ and $\Delta_b$, as well as the rates
$\gamma_{\alpha}$ and $ \Gamma_{\beta}$, which describe the tunneling between the active sites and the reservoirs, are small
enough compared to a parameter $\sqrt{\lambda T}$ which defines a characteristic energy scale of the noise operator
$\xi=\xi_a+\xi_b$, with a {\it combined reorganization energy} $$\lambda =\lambda_a+\lambda_b.$$ Then, all calculations can be
done with an accuracy up to second order in the F\"orster energy, $|V_F|^2,$ and up to second order for the direct tunneling
rates, $|\Delta_a|^2$ and $|\Delta_b|^2.$ The electron (proton) current  consists of two components, $i_{\alpha F} ~(I_{\beta
F}),$ related to the F\"orster process, and $i_{\alpha,\, {\rm dir}} ~(I_{\beta,\, {\rm dir}}),$ describing the contributions of
direct tunneling to the electron (proton) flow. The F\"orster components of the electron and proton currents are given by the same
expression (up to the total sign):
\begin{equation}
i_{RF} = -i_{LF} = I_{PF} = - I_{NF} =   i V_F^* \langle e^{-i\xi} \rho_8^7 \rangle - i V_F \langle \rho_7^8 ~e^{i\xi}\rangle.
\label{Cur4}
\end{equation}
 The direct electron (proton) current $i_{R,\, {\rm dir}}\
(I_{N,\, {\rm dir}}) $ is proportional to the tunneling rate $\Delta_a \;(\Delta_b) :$
\begin{eqnarray}
i_{R,\, {\rm dir}} = - i_{L,\, {\rm dir}} = i \Delta_a^* \langle  ( \rho_2^3  + \rho_6^8  + \rho_7^9  + \rho_{14}^{15} )e^{i\xi_a} \rangle + h.c. \nonumber\\
I_{N,\, {\rm dir}} = - I_{P,\, {\rm dir}} = i \Delta_b^* \langle e^{-i\xi_b}( \rho_4^5  + \rho_6^7  + \rho_8^9  + \rho_{11}^{12} )
\rangle + h.c. \label{Cur5}
\end{eqnarray}

\subsection{Calculation of the F\"orster current}

To calculate the F\"orster component of the current up to second
order in the energy $V_F$, we derive the Heisenberg equation for
the operator $ \rho_7^8 $ neglecting the coupling to the
reservoirs and the direct tunneling:
\begin{equation}
i\frac{d}{d t} \rho_7^8 = \delta ~\rho_7^8 + V_F^* e^{-i \xi} ~( \rho_{7} - \rho_{8} ), \label{Rho78}
\end{equation}
where $\delta$ is the  detuning between the electron and proton energy levels,
\begin{equation}
\delta \;=\; \varepsilon_8 - \varepsilon_7 \;=\; \epsilon_2 - \epsilon_1 - E_2 + E_1 - u_{21} + u_{12}. \label{Detun}
\end{equation}
 The solution of Eq.~(\ref{Rho78}),
\begin{equation}
\rho_7^8(t) = - i V_F^* \int_{-\infty}^t dt_1\; e^{-i \delta (t-t_1)} e^{-i\xi(t_1)} [ \rho_{7}(t_1) - \rho_{8}(t_1)],
\label{Rho78F}
\end{equation}
should be substituted in Eq.~(\ref{Cur4}) for the current $i_{RF}$,
\begin{equation}
i_{RF} = - |V_F|^2 \int_{-\infty}^t dt_1 \; e^{-i \delta (t-t_1)} \langle e^{-i\xi(t_1)} e^{i\xi(t)}\rangle \langle\rho_7 -
\rho_8\rangle (t_1) + h.c. \label{CurRF}
\end{equation}
Here, we separate the averaging of the environment phases $\xi =
\xi_a + \xi_b $ from the operators of the electron-proton
subsystem. For independent electron and proton environments, when
$$ \langle e^{-i\xi(t_1)} \, e^{i\xi(t)}\rangle = \langle e^{-i\xi_a(t_1)}\, e^{i\xi_a(t)}\rangle \langle e^{-i\xi_b(t_1)} \,
e^{i\xi_b(t)}\rangle,$$ we can also calculate the electron and proton functionals separately. In particular, for the electronic
environment characterized by the operator $\xi_a = \sum_j x_{j0}\,p_j$ (from here on $\hbar = 1$) we obtain the relation
$$ \exp\{-i\xi_a(t)\} \, \exp\{i\xi_a(t_1)\} = \exp\{ -i[\xi_a(t)-\xi_a(t_1)]\} \, \exp\{(1/2)[\xi_a(t),\xi_a(t_1)]_-\}, $$
where the commutator,  $$(1/2)[\xi_a(t),\xi_a(t_1)]_- = -i \sum_j m_j \omega_j x_{j0}^2 \sin \omega_j(t-t_1), $$ is determined
using the free-evolving oscillator operators,
$$ x_j(t) = x_j(t_1) \cos \omega_j(t-t_1) + \frac{p_j}{m_j\omega_j} \sin \omega_j(t-t_1), $$
$$ p_j(t) = p_j(t_1) \cos \omega_j(t-t_1) - m_j\omega_j x_j \sin \omega_j(t-t_1). $$
For the Gaussian statistics of the system of independent oscillators, the characteristic functional has the form
$$ \langle \exp\{-i[\xi_a(t)-\xi_a(t_1)] \} \rangle = \exp\{ - \langle \xi_a^2\rangle +
\frac{1}{2}\langle [ \xi_a(t), \xi_a(t_1)]_+\rangle \},$$ with
$$ \frac{1}{2}\langle [ \xi_a(t), \xi_a(t_1)]_+\rangle = \sum_j x_{j0}^2
\frac{1}{2}\langle [ p_j(t),p_j(t_1)]_+\rangle = \sum_j \langle
p_j^2\rangle x_{j0}^2 \cos \omega_j(t-t_1). $$ Taking into account
the expression for the equilibrium dispersion of the
$j$th-oscillator momentum, $ \langle p_J^2\rangle =
(m_j\omega_j/2)\coth(\omega_j/2T),$ we obtain the well-known
expression \cite{Krish01}   for the functional  $ \langle
e^{-i\xi_a(t)} e^{i\xi_a(t_1)} \rangle$:
\begin{equation}
 \langle \exp\{-i\xi_a(t)\} \exp\{i\xi_a(t_1)\} \rangle =  \exp\{-i W_{1a}(t)\} \exp\{-W_{2a}(t)\}, \label{Func1}
\end{equation}
where
\begin{equation}
W_{1a}(t) =\sum_j \frac{m_j\omega_jx_{j0}^2}{2} \sin \omega_j t \,=\, \int_0^{\infty} d\omega \frac{J_a(\omega)}{\omega^2}
\sin\omega t,\label{W1a}
\end{equation}
and
\begin{equation}
W_{2a}(t) =\sum_j \frac{m_j\omega_jx_{j0}^2}{2} \coth\left(\frac{\omega_j}{2T}\right) (1- \cos \omega_j t)  \;=\; \int_0^{\infty}
d\omega \frac{J_a(\omega)}{\omega^2}\coth\left(\frac{\omega}{2T}\right) ( 1 - \cos \omega t ). \label{W2a}
\end{equation}
Similar relations between $W_{1b}(t), W_{2b}(t)$ and the spectral
function $J_b(\omega)$ take place for the proton dissipative
environment. Notice that for this model, the effects of the
electrons and protons on the environments are disregarded. In the
semiclassical approximation $ (T \gg \omega)$ and for slow enough
fluctuations of the environments $(\omega t \ll 1)$, the functions
$ W_{1a}(t), W_{2a}(t)$ have simple forms $$W_{1a}(t) = \lambda_a
t,\ \  W_{2a}(t) = \lambda_a T t^2.$$ Thus, we have
\begin{equation}
 \langle \exp\{-i\xi_a(t)\} \exp\{i\xi_a(t_1)\} \rangle =  \exp\{-i \lambda_a (t-t_1) \} \exp\{-\lambda_a T
 (t-t_1)^2\}. \label{Func2}
\end{equation}
The total characteristic functional involved in Eq.~(\ref{CurRF})
for the F\"orster current, $ \langle e^{-i\xi(t)} e^{i\xi(t_1)}
\rangle = e^{-i \lambda (t-t_1) } e^{-\lambda T  (t-t_1)^2},$ has
an effective correlation time $(\hbar =1)$, $$\tau_{c} =
\frac{1}{\sqrt{\lambda T}}, $$ which is determined by the {\it
combined} electron-proton reorganization energy, $\lambda =
\lambda_a + \lambda_b.$ At strong enough electron-proton couplings
to the surroundings, the correlation time $\tau_c$ is much shorter
than the time scale of the probabilities $\rho_n$, so that in
Eq.~(\ref{CurRF}) we can put $\langle\rho_7-\rho_8\rangle(t_1)
\simeq \langle \rho_7-\rho_8\rangle(t).$ It allows us to obtain a
simple expression for the F\"orster current:
\begin{equation}
 i_{RF} = -i_{LF} = I_{PF} = - I_{NF} = \kappa \langle \rho_8 - \rho_7\rangle, \label{CurRFin}
\end{equation}
 where $\kappa$ looks like the well-known semiclassical
Marcus rate \cite{Krish01,MarSut85},
\begin{equation} \kappa = \sqrt{\frac{\pi}{\lambda T}} |V_F|^2 \exp\left[ - \frac{(\delta - \lambda )^2}{4\lambda T}\right],
\label{kappa1}
\end{equation}
but with the only difference that instead of the reaction free
energy of a proton pumping step, $ \Delta G \sim E_2-E_1 \sim
\epsilon_2 - \epsilon_1, $ here we have the {\it electron-proton
detuning}, $$\delta = \epsilon_2 - \epsilon_1 - E_2 + E_1 - u_{21}
+ u_{12},$$ which is much smaller and can be even zero for the
case of an exact electron-proton resonance. Near these {\it
resonant } conditions, when $\delta = \lambda$, the proton pump
should be most effective.

\subsection{Direct currents}

Similar calculations (not shown here) demonstrate that the direct electron (proton) current, Eq.~(\ref{Cur5}), is proportional to
the standard non-resonant Marcus rate $k_a \; (k_b)$:
\begin{eqnarray}
i_{R,\, {\rm dir}} = - i_{L,\, {\rm dir}} = k_a \langle \rho_3 + \rho_8 + \rho_9 + \rho_{15} - \rho_2 - \rho_6 - \rho_7 - \rho_{14} \rangle, \nonumber\\
I_{N,\, {\rm dir}} = - I_{P,\, {\rm dir}} = k_b \langle \rho_5 + \rho_7 + \rho_9 + \rho_{12} - \rho_4 - \rho_6 - \rho_8 -
\rho_{11} \rangle, \label{Curdir1}
\end{eqnarray}
where
\begin{eqnarray}
\kappa_a = \sqrt{\frac{\pi}{\lambda_a T}} |\Delta_a|^2 \exp\left[ - \; \frac{(\epsilon_2 - \epsilon_1 - \lambda_a )^2}{4\lambda_a
T}\right], \nonumber\\
\kappa_b = \sqrt{\frac{\pi}{\lambda_b T}} |\Delta_b|^2 \exp\left[ - \; \frac{(E_2 - E_1 - \lambda_b )^2}{4\lambda_b T}\right].
\label{kapAB}
\end{eqnarray}
The processes of {\it direct} electron and proton tunnelings lead
to the {\it downhill transfer} of protons, {\it discharging} the
proton battery. However, this process is significantly suppressed
when the separation of the proton energy levels is much higher
than the reorganization energy $\lambda_b.$

\section{Density matrix}

The electron and proton currents, Eqs.~(\ref{CurRFin}) and
(\ref{Curdir1}), are determined by the diagonal elements of the
density matrix of the electron-proton system $\langle \rho_m
\rangle $ over the eigenstates, Eq.~(\ref{basis1}), of the
Hamiltonian, Eq.~(\ref{H01}). To obtain the diagonal elements of
the density matrix, we write the Heisenberg equation for the
operators $\rho_m$ taking into account the basis Hamiltonian $H_0
= \sum_n \varepsilon_n \rho_n, $ complemented by terms which are
responsible for: (i) the F\"orster process $H_F$, (ii) the direct
tunneling events between the active sites $H_{\rm dir}$, and
 (iii) the tunneling coupling between the reservoirs and the active sites $H_{\rm tun}$,
$$ i\dot{\rho}_m \  = \ [H,\rho_m]_- \ = \  [\rho_m,H_F]_-  + [\rho_m,H_{\rm dir}]_- +  [\rho_m,H_{\rm tun}]_- \ .
$$
With the tunneling Hamiltonian, Eq.~(\ref{Htun}), where the
electron and proton operators are represented as expansions, $$
a_{\sigma} = \sum_{mn} a_{\sigma;mn}\,\rho_m^n, \ \ \  b_{\sigma}
= \sum_{mn} b_{\sigma;mn}\,\rho_m^n $$ (see Eq.~(\ref{ab12}) ), we
obtain the contribution of the two pairs of reservoirs to the
evolution of the operator $\rho_m$ as
\begin{eqnarray}
[\rho_m,H_{\rm tun}]_- = - \sum t_{kR} ~e^{-i\xi_a/2}~ c_{kR}^+ ~( a_{1;mn} \, \rho_m^n - a_{1;nm} \, \rho_n^m ) - \nonumber\\
\sum t_{kL}  ~c_{kL}^+ ~( a_{1;mn}\, \rho_m^n - a_{1;nm}\, \rho_n^m )\,e^{i\xi_a/2} - \nonumber\\
\sum T_{qN} ~ d_{qN}^+ ~ ( b_{1;mn}\, \rho_m^n - b_{1;nm} \, \rho_n^m )\,e^{i\xi_b/2} - \nonumber\\
\sum T_{qP}~ e^{-i\xi_b/2}~ d_{qP}^+ ~ ( b_{2;mn} \,\rho_m^n - b_{2;nm} \,\rho_n^m ) - \{ h.c.\}, \label{RhoTun1}
\end{eqnarray}
Substituting Eq.~(\ref{cdReact}) for the leads reactions, and averaging over the Fermi distributions of electrons and protons in
the leads and over the fluctuations of the environments, we obtain the contribution of leads to the master equation for the
probabilities $\langle \rho_m\rangle:$
\begin{eqnarray}
\langle [\rho_m,H_{\rm tun}]_-\rangle = i \sum_n ( \gamma_{mn}^{tun} \langle \rho_n\rangle - \gamma_{nm}^{tun} \langle
\rho_m\rangle ), \label{RhoTun2}
\end{eqnarray}
with the relaxation matrix
\begin{eqnarray}
\gamma_{mn}^{\rm tun} = \gamma_R \{ |a_{1;mn}|^2 [ 1 - f_R(\omega_{nm}) ] + |a_{1;nm}|^2 f_R(\omega_{mn}) \} + \nonumber\\
\gamma_L \{ |a_{2;mn}|^2 [ 1 - f_L(\omega_{nm}) ] + |a_{2;nm}|^2 f_L(\omega_{mn}) \} + \nonumber\\
\Gamma_N \{ |b_{1;mn}|^2 [ 1 - F_N(\omega_{nm}) ] + |b_{1;nm}|^2 F_N(\omega_{mn}) \} + \nonumber\\
\Gamma_P \{ |b_{2;mn}|^2 [ 1 - F_P(\omega_{nm}) ] + |b_{2;nm}|^2 F_P(\omega_{mn}) \} . \label{gamTun1}
\end{eqnarray}
The products of free reservoir operators, such as $c_{k\alpha}^{(0)}(t)$, and an arbitrary Fermi operator of electrons, ${\cal
Z}_F$, can be calculated using the formula
\begin{equation}
\langle {\cal Z}_F(t) c_{k\alpha}^{(0)}(t)\rangle = - i t_{k\alpha\sigma}\int dt_1\; \langle c_{k\alpha}^{(0)+}(t_1)
c_{k\alpha}^{(0)}(t)\rangle \,\langle [{\cal Z}_F(t),a_{\sigma}(t_1)]_+ \rangle\,\theta (t-t_1). \label{FermiCor}
\end{equation}
Similar formulas can be employed for the proton component. The F\"orster process contributes to the evolution of two components of
the density matrix, $ \rho_7$ and $\rho_8$,
\begin{equation}
[\rho_7,H_{F}]_- = - [\rho_8,H_{F}]_- = V_F ~\rho_7^8 ~e^{i\xi} -  V_F^*~ e^{-i\xi}~\rho_8^7. \label{Rho7F1}
\end{equation}
Due to the  weakness of the tunneling processes, we disregard the overlap of the different tunneling mechanisms in the master
equation for the distribution $\langle \rho_m\rangle.$ Substituting Eq.~(\ref{Rho78F}) for the operator $\rho_7^8$ and its
conjugate jointly with Eq.~(\ref{Func2}) for the characteristic functional of the environments, we obtain the contribution of the
F\"orster process to the master equation as
\begin{equation}
\langle [\rho_7,H_{F}]_-\rangle  = - \langle [\rho_8,H_{F}]_-\rangle = i \kappa ( \langle \rho_8\rangle - \langle\rho_7\rangle ),
\label{Rho7F2}
\end{equation}
where $\kappa$ is the resonant Marcus rate, Eq.~(\ref{kappa1}). In a similar way, we determine that the direct tunneling between
the active sites contributes to the equations for the following probabilities:
$$
\langle [\rho_2,H_{\rm dir}]_-\rangle = - \langle [\rho_3,H_{\rm dir}]_-\rangle  = i \kappa_a ( \langle \rho_3\rangle -
\langle\rho_2\rangle ), $$
$$ \langle [\rho_4,H_{\rm dir}]_- \rangle = - \langle [\rho_5,H_{\rm dir}]_-\rangle  = i \kappa_b ( \langle \rho_5\rangle -
\langle\rho_4\rangle ), $$
$$\langle [\rho_6,H_{\rm dir}]_- \rangle  = i \kappa_a ( \langle \rho_8\rangle - \langle\rho_6\rangle ) + i
\kappa_b ( \langle \rho_7\rangle - \langle\rho_6\rangle ),  $$
$$ [\rho_7,H_{\rm dir}]_-  = i \kappa_a ( \langle \rho_9\rangle -
\langle\rho_7\rangle ) - i \kappa_b ( \langle \rho_7\rangle - \langle\rho_6\rangle ),  $$ $$  [\rho_8,H_{\rm dir}]_-  = - i
\kappa_a ( \langle \rho_8\rangle - \langle\rho_6\rangle ) + i \kappa_b ( \langle \rho_9\rangle - \langle\rho_8\rangle ),  $$ $$
[\rho_9,H_{\rm dir}]_- = -i \kappa_a ( \langle \rho_9\rangle - \langle\rho_7\rangle ) - i \kappa_b ( \langle \rho_9\rangle -
\langle\rho_8\rangle ), $$ $$ [\rho_{11},H_{\rm dir}]_- = - [\rho_{12},H_{\rm dir}]_- = i \kappa_b ( \langle \rho_{12}\rangle -
\langle\rho_{11}\rangle ), $$ $$  [\rho_{14},H_{\rm dir}]_- = - [\rho_{15},H_{\rm dir}]_- = i \kappa_a ( \langle \rho_{15}\rangle
- \langle\rho_{14}\rangle ),
$$
where $k_a$ and $k_b$ are the non-resonant Marcus rates given by
Eq.~(\ref{kapAB}). Combining all contributions, we obtain the
following master equation for the probabilities $\langle
\rho_m\rangle$:
\begin{equation}
\langle \dot{\rho}_m\rangle + \gamma_m \langle \rho_m\rangle = \sum_n \gamma_{mn} \langle \rho_n\rangle, \label{RhoEq1}
\end{equation}
with the relaxation rates $\gamma_m = \sum_n \gamma_{nm},$   where $ \gamma_{mn} = \gamma_{mn}^{\rm tun} $ given by
Eq.~(\ref{gamTun1}) for all matrix elements except
\begin{eqnarray}
\gamma_{2,3} = \gamma_{2,3}^{\rm tun}+k_a; \gamma_{3,2} = \gamma_{3,2}^{\rm tun}+k_a;
\gamma_{4,5} = \gamma_{4,5}^{\rm tun}+k_b; \gamma_{5,4} = \gamma_{5,4}^{\rm tun}+k_b; \nonumber\\
\gamma_{6,7} = \gamma_{6,7}^{\rm tun}+k_b; \gamma_{7,6} = \gamma_{7,6}^{\rm tun}+k_b;
\gamma_{6,8} = \gamma_{6,8}^{\rm tun}+k_a; \gamma_{8,6} = \gamma_{8,6}^{\rm tun}+k_a; \nonumber\\
\gamma_{7,8} = \gamma_{7,8}^{\rm tun}+\kappa; \gamma_{8,7} = \gamma_{8,7}^{\rm tun}+\kappa;
\gamma_{7,9} = \gamma_{7,9}^{\rm tun}+k_a; \gamma_{9,7} = \gamma_{9,7}^{\rm tun}+k_a; \nonumber\\
\gamma_{8,9} = \gamma_{8,9}^{\rm tun}+k_b; \gamma_{9,8} = \gamma_{9,8}^{\rm tun}+k_b;
\gamma_{11,12} = \gamma_{11,12}^{\rm tun}+k_b; \gamma_{12,11} = \gamma_{12,11}^{\rm tun}+k_b; \nonumber\\
\gamma_{14,15} = \gamma_{14,15}^{\rm tun}+k_a; \gamma_{15,14} = \gamma_{15,14}^{\rm tun}+k_a. \label{gammaMN}
\end{eqnarray}

It should be noted that the {\it key ingredient} of the proposed
model is the {\it resonant F\"orster exchange of energy} between
electrons and protons. This process takes place in a time interval
$$\tau_F= \frac{1}{2\kappa},$$ where $\kappa$ is the resonant
Marcus rate Eq.~(\ref{kappa1}), as follows from the solution of
the rate equations, $ \langle \dot{\rho}_7 \rangle = - \kappa
\langle \rho_7-\rho_8\rangle = -\langle \dot{\rho}_8 \rangle,$
derived in the absence of the leads. If our system is initially in
the state $\vert 8\rangle$ with the excited electron and with the
proton in the ground state, then, the probability to be in the
state $\vert 7\rangle$, where the proton is on the upper level and
the electron in the ground state, is given by the formula $$
\rho_{7} (t) = ( 1 - e^{-2\kappa t})/2.$$ After a lapse of time
scale $ \tau_F$, the proton goes to the excited state with
probability $1/2$.

\section{Results and discussion}

The steady-state version of Eq.~(\ref{RhoEq1}),
\begin{equation}
\sum_n \gamma_{nm} \; \langle \rho_m\rangle = \sum_n \gamma_{mn}\; \langle \rho_n\rangle , \label{Steady}
\end{equation}
 $(m,n=1,..16)$, has been solved numerically
jointly with the normalization condition $\sum_m \rho_m = 1,$ with
subsequent calculations of the electron and proton currents
through the system, Eqs.~(\ref{CurRFin}),(\ref{Curdir1}), and
populations of all active sites, $ \langle n_{\sigma}\rangle$ and
$\langle N_{\sigma}\rangle.$ To obtain numerical values, we assume
that the electron potential well, presumably attached to the
binuclear center, contains two active electron sites and has a
radius $r_0$ of about $0.1$ nm. The proton potential well with a
radius $R_0 \sim 0.01$ nm can be located at the pump center X at a
distance $R \sim 1$ nm from the electron sites. Thus, in a medium
with a dielectric constant $\epsilon_r$\;=\;3 (dry protein), the
F\"orster constant in Eq.~(\ref{VF1}) has a $V_F \sim 1$ meV.
Taking into account renormalization effects for the direct Coulomb
coupling between electrons and protons, we choose $$u_{11}\simeq
u_{12}\simeq u_{21}\simeq u_{22} = 400 \;{\rm meV}$$ which is
close to the energy of the Coulomb interaction, $u\simeq 480$ meV,
of two charges located a distance $R\,\simeq \,1$ nm apart. The
on-site Coulomb repulsion energies,$u_e$ and $u_p$, are estimated
as
$$u_e\simeq u_p \simeq 4000 \; {\rm meV}, $$ which is enough to
avoid the double-occupation of the active sites. For the rates of
the possible direct electron and proton transitions between the
active sites, we take the values $\Delta_a = 1$ meV and $\Delta_b
= 0.1$ meV, respectively. The tunneling couplings of the electrons
to the leads are $\Gamma_L = \Gamma_R =$ 0.85 meV,  and the proton
rates are $\Gamma_N = \Gamma_P = 0.1$
 meV. For the optimal efficiency of the pump, we choose the energy levels of the electron and proton active sites as
$$\epsilon_1 =  100 \;{\rm meV}, \; \epsilon_2 =  600 \;{\rm meV}$$ and $$E_1 =  350 \; {\rm meV}, \; E_2 \simeq  850 \; {\rm meV},$$
so that the difference between the electron energy levels
$\epsilon_2$ and $\epsilon_1,$ corresponds to the realistic drop
of the COX redox potential \cite{Wik04,Hos06}, and it is in
resonance with the separation of proton levels $$\epsilon_2 -
\epsilon_1 = E_2 - E_1 =
 500 \;{\rm meV}. $$

 We consider here intermediate values of the reorganization energies, $$\lambda_a \simeq \lambda_b \simeq  3 \ {\rm meV},\;
\lambda \simeq 6 \ {\rm meV},$$ which are higher than the F\"orster constant $V_F$ and all other tunneling rates. Then the Marcus
constants related to the direct tunneling, $k_a,k_b$, Eq.~(\ref{kapAB}), are negligibly small $(\sim 10^{-100}$ meV/$\hbar$);
however, the F\"orster rate, Eq.~(\ref{kappa1}), is quite pronounced, $\kappa \simeq  0.1 \ {\rm meV}/\hbar \simeq 150 \ {\rm
ns}^{-1} $. The rates $\kappa_a,\kappa_b,$ and $\kappa$ can be measured in the units of ${\rm meV}/\hbar$ or in the inverse
nanoseconds (ns): $ 1 \ {\rm meV}/\hbar \simeq 1500 \ {\rm ns}^{-1}.$ The real values of the reorganization energies
$\lambda_a,\lambda_b$ are not known yet for the enzyme cytochrome $c$ oxidase, although it is expected that they are of order or
higher than 100 meV \cite{Ols07, Krish01}. These numbers can be estimated from measurements of the temperature dependence of the
Marcus rates $\kappa_a,\kappa_b$ (\ref{kapAB}) for the transitions between the active electron and proton sites.

It should be noted that at the reorganization energies $\lambda_a,
\lambda_b\simeq $ 100 meV, and at the physiological temperature, $
T=36.6^{\circ}$C, direct tunneling processes are also
significantly suppressed, $$\kappa_a \sim 10^{-5}\  {\rm ns}^{-1},
 \ \  \kappa_b \sim 10^{-15} \ {\rm ns}^{-1}.$$ However, the F\"orster mechanism of energy
  transfer survives near the electron-proton resonance with the rate $\kappa \sim 30 \ {\rm ns}^{-1} $.
This means that even for the case of strong coupling to the
dissipative environments, the pure electron-proton F\"orster
exchange (with no leads) occurs over the time scale $$ \tau_F =
1/(2\kappa) \sim 20 \;{\rm ps}.$$

In the following, all contributions of the direct tunneling are
disregarded,  so that the total particle current is exclusively
determined by the F\"orster component, Eq.~(\ref{CurRFin}), and
the electron flow from the left reservoir to the right one, $i_R$,
is exactly equal to the particle current of protons, $$ I_{P} = -
I_{N} = i_R,$$ flowing from the negative side to the positive side
of the membrane against the concentration gradient. In other
words, one proton is pumped through the membrane per each electron
transferred to the oxygen molecule ${\rm O}_2$ that can play the
role of our right electron reservoir, consistent with experimental
observations of Refs.~\cite{Brand06,Belev07,Bloch04}. It should be
mentioned that in the present model, we do not consider  substrate
protons, which are also taken from the negative side of the
membrane to form the water molecules.

\subsection{Pumping effects}

Here, the positive direction of the current is defined to be from
the higher chemical potential to the lower chemical potential. The
electrochemical potential of the left electron lead, $\mu_L,$ is
chosen to be higher than the potential of the right lead at the
positive voltage $V_e$: $$\mu_L = V_e, \ \ \  \mu_R = 0,$$ whereas
for the protons the chemical potential of the positive side of the
membrane, $\mu_P,$ exceeds the potential of the negative side at
the positive voltage $V_p$: $$\mu_P = V_p, \ \ \ \mu_N = 0.$$
Notice that throughout the paper the ``voltages" $V_e,V_p$
incorporate the absolute value of the electron charge and are
measured in meV. When the electron voltage is positive, $V_e>0,$
the electron particle current $i_R$, Eq.~(\ref{CurDef}), should be
positive because the electron concentration of the right lead
increases. At normal conditions, the protons should also flow from
the positive side of the membrane (having a higher chemical
potential at $V_p>0$) to the negative side, so that the population
of protons on the negative side should grow, that corresponds to a
positive particle current $I_N.$

In Fig.~2, we present the numerical solution for the dependence of the proton current $I_N$ on the electron ($V_e$) and proton
($V_p$) voltages at the physiological temperature $ T=36.6^{\circ}$C,  with $E_2=850$ meV. The particle current is measured here
in the inverse nanoseconds, ${\rm ns}^{-1}$, so that, for example, the value $I_N = - 1 \ {\rm ns}^{-1}$ corresponds to the
transfer of one proton per one nanosecond from the negative side of the membrane to the positive side. It is evident from Fig.~2
that the uphill proton current (corresponding to negative values of $I_N$) starts at electron voltages exceeding a threshold value
$V_{e0} = 550$ meV provided that the proton voltage build-up is less than $450$ meV. At these voltages, the states $$|7\rangle =
a_1^+b_2^+|{\rm Vac}\rangle \ \ {\rm and}\ \ |8\rangle = a_2^+b_1^+|{\rm Vac}\rangle$$ participating in the F\"orster transfer
(see Eq.~(\ref{CurRFin})) and having energies $\sim 550 $ meV begin to be populated. It is of interest that at lower voltages the
state $|6\rangle = a_1^+ b_1^+|{\rm Vac}\rangle$ containing an electron in the state $1_e$ with energy $\epsilon_1=100$ meV and a
proton in the state $1_p$, having an energy $E_1=350$ meV, is partially populated. Here, the electron-proton Coulomb attraction,
$u_{11}=-400$ meV, comes into play, lowering the total energy to the value $\varepsilon_6 = 50$ meV.

For the chosen parameters, the particle current $I_N$ saturates at electron voltages higher than 700 meV with the value
corresponding to the translocation of 30 protons in one nanosecond. It shows the efficiency of the F\"orster pumping mechanism,
although the real rate for the proton transfer through the D-pathway (see Ref.~\cite{Brand06}) is much less: $\sim 10^3$--$10^4 $
protons per second. This pumping rate can be obtained in the framework of our model if we significantly decrease the tunneling
couplings between the active sites and the electron and proton reservoirs: $\Gamma_L\sim \Gamma_R \sim 10^{-7} \ {\rm meV}, \ \
\Gamma_N\sim \Gamma_P \sim 10^{-8} \ {\rm meV}.$ It has no effect on the main features of the present model, and, in the
following, we return to the  case of the fast electron and proton delivery to the active sites.

If the electron voltage is low enough, $V_e < 300 $ meV, but the proton voltage is high, $V_p > 500 $ meV, the proton flow
reverses its direction, so that the protons move along the concentration gradient from the positive side of the membrane to the
mitochondria interior. The downhill flow of the protons is especially significant when the proton voltage exceeds the value of 850
meV. However, even at high proton voltages, the discharge of the mitochondrion battery can be prevented by applying the electron
potential above the threshold $V_{e0} = 550$ mV. We emphasize that, within this model, we do not need any additional gates to
inhibit the translocation of protons back to the negatively-charged interior, although the pump can work in the reverse regime.
The optimal value for the proton voltage build-up, $V_p=250$ meV, correlates well with experimental data for the proton-motive
force of about 200--250 meV \cite{Wik04,Brand06,Papa04}.

The resonant character of the F\"orster energy transfer is
demonstrated in Fig.~3 where we plot a dependence of the proton
current $I_N$ on the variation of the higher energy level of the
protons, $E_2$, at several temperatures $T$ measured in degrees
Celsius. It is evident that the current $I_N$ has the maximum
absolute value at the energy $$E_2 = \epsilon_2-\epsilon_1 +E_1 -
\lambda = 844 \ {\rm meV},$$ which is slightly shifted from its
resonance value $E_2=850$ meV in accordance with the maximum of
the Marcus constant $\kappa$, Eq.~(\ref{kappa1}).

In Fig.~4 we present the temperature dependence of the uphill
proton current near the optimal point $$V_e\,=\,700 \ {\rm meV},\;
V_p\,=\,250 \ {\rm meV},\; E_2 = 850 \  {\rm meV}.$$ It is clear
that the proton pumping peaks at temperatures between $0
^{\circ}$C and $100^{\circ}$C with a strong decrease when the
environment is colder than the water freezing point $0 ^{\circ}$C.
However, the effect survives much better at high temperatures.
Curiously, for the parameters used the uphill proton current has a
maximum at temperatures about that of the human body $(36.6
^{\circ}$C).

\section{Conclusions}

In conclusion, we proposed and analyzed quantitatively a simple
nano-electronic and nano-protonic model reflecting the main
features of the electron-driven proton pump in the enzyme {\it
cytochrome c oxidase}. We analyzed quantum-mechanical Hamiltonians
for this system taking into account tunneling couplings of
electrons and protons to their corresponding reservoirs and
dissipative environments, as well as the electron-proton Coulomb
interaction, including the resonant F\"orster term. Applying
methods of condensed matter physics, we obtained expressions for
the electron and proton currents as well as the  equations of
motion for the density matrix of the system. These equations were
solved numerically, and we demonstrated that the resonant
F\"orster energy exchange between electrons and protons can lead
to the proton transfer from the region with smaller proton
concentration to the region with larger proton concentration,
thereby achieving a proton pump. The dependence of this phenomenon
on temperature and the system parameters were studied and we
showed that the proton pump works with maximum efficiency near
physiological temperatures and at electron and proton voltage
build-ups related to their values for living cells.

{\bf Acknowledgements}

 This work was supported in part by the National Security Agency, Laboratory of Physical Sciences, Army
Research Office, National Science Foundation grant No. EIA-0130383, and JSPS CTC Program. L.M. is partially supported by the NSF
NIRT, grant ECS-0609146.

\begin{figure}
\includegraphics[width=18.0cm]{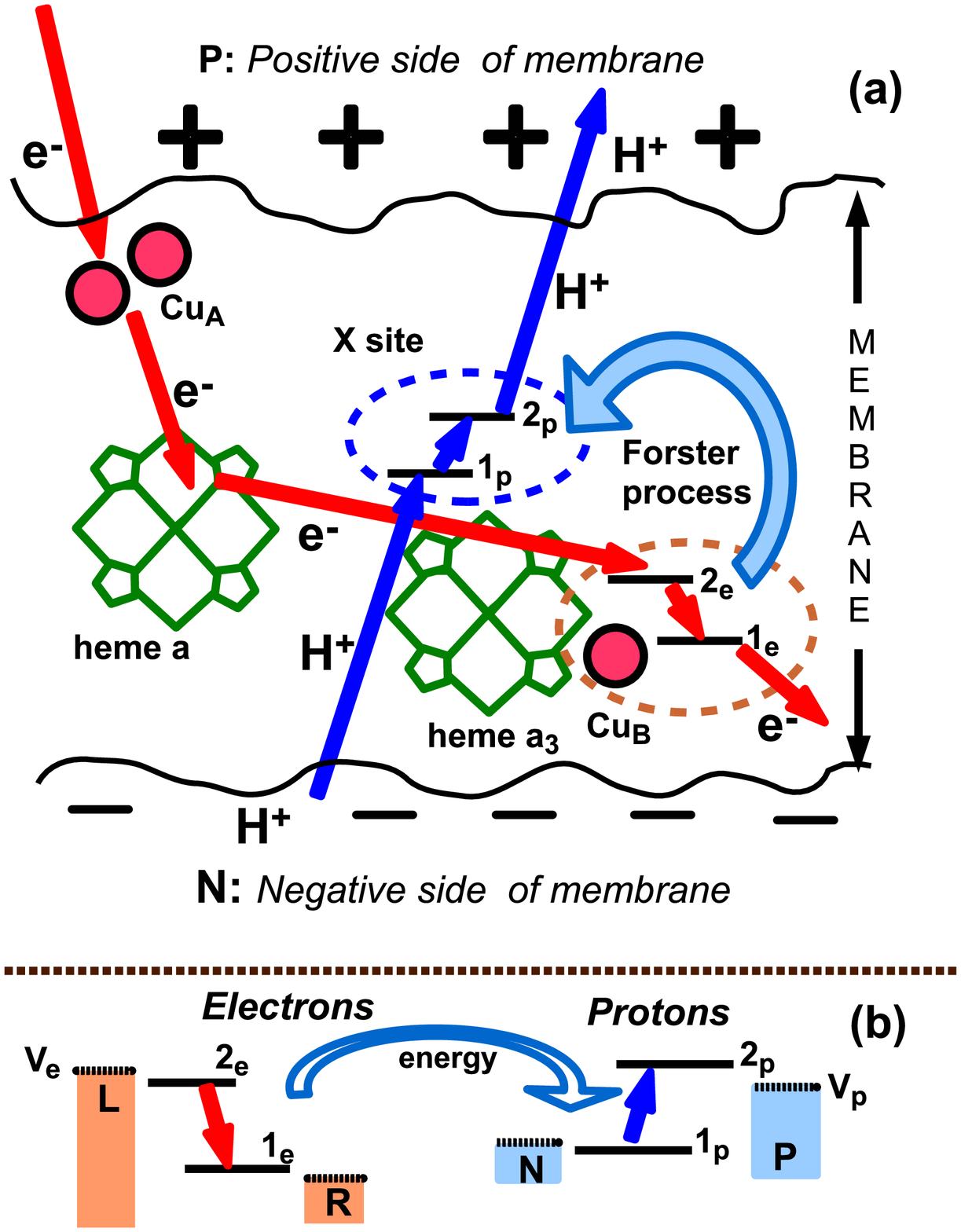}
\vspace*{-4cm} \caption{ (Color online) (a) Schematic diagram of the electron and proton pathways in cytochrome  $c$ oxidase with
suggested locations for the active electron and proton sites. (b) Schematic energy diagram of the simultaneous electron and proton
transport.}
\end{figure}

\begin{figure}
\includegraphics[width=18.0cm]{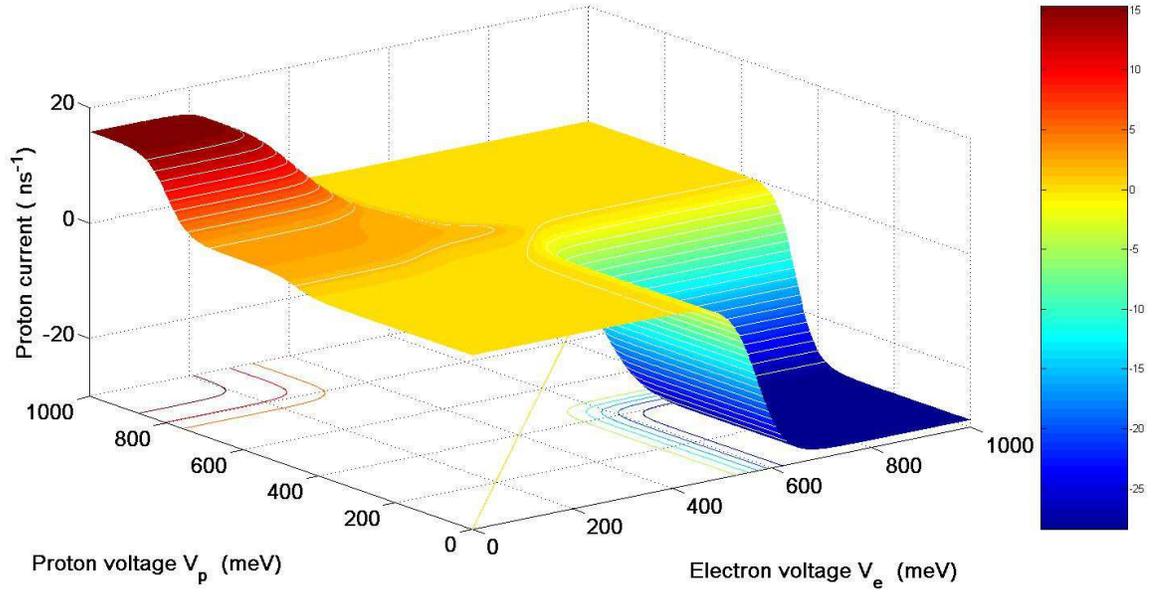} 
\vspace*{0cm} \caption{(Color online) Proton current $I_N$ (a number of protons transferred through the membrane in one
nanosecond) as a function of the electron $(V_e)$ and proton $(V_p)$ voltage build-ups at the physiological temperature
$T=36.6^{\circ}$C and at the resonant condition, $E_2 = 850$ meV. Notice that the absolute value of the electron charge, $|e|$, is
included into the definitions of voltages $V_e,~V_p$, which are measured here in meV. }
\end{figure}

\begin{figure}
\includegraphics[width=18.0cm]{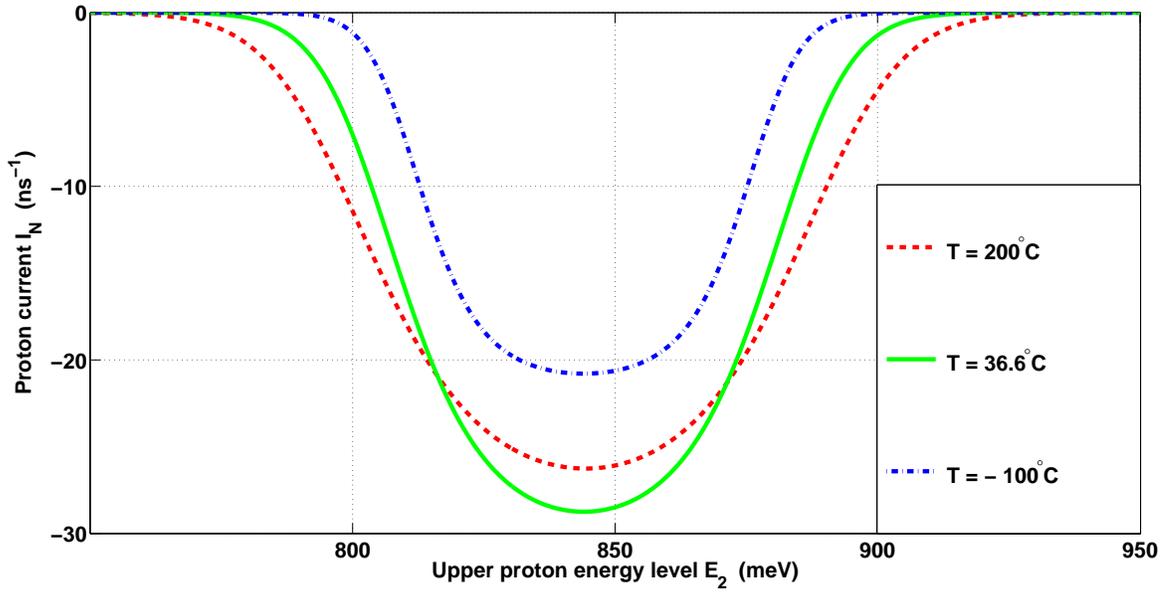}
\vspace*{0cm} \caption{(Color online) Dependence of the proton current $I_N$ on the resonant conditions (a variation of the upper
proton energy level $E_2$) at different temperatures, for optimal values of the electron and proton voltages: $V_e=700$ meV,
$V_p=250$ meV.}
\end{figure}

\begin{figure}
\includegraphics[width=18.0cm]{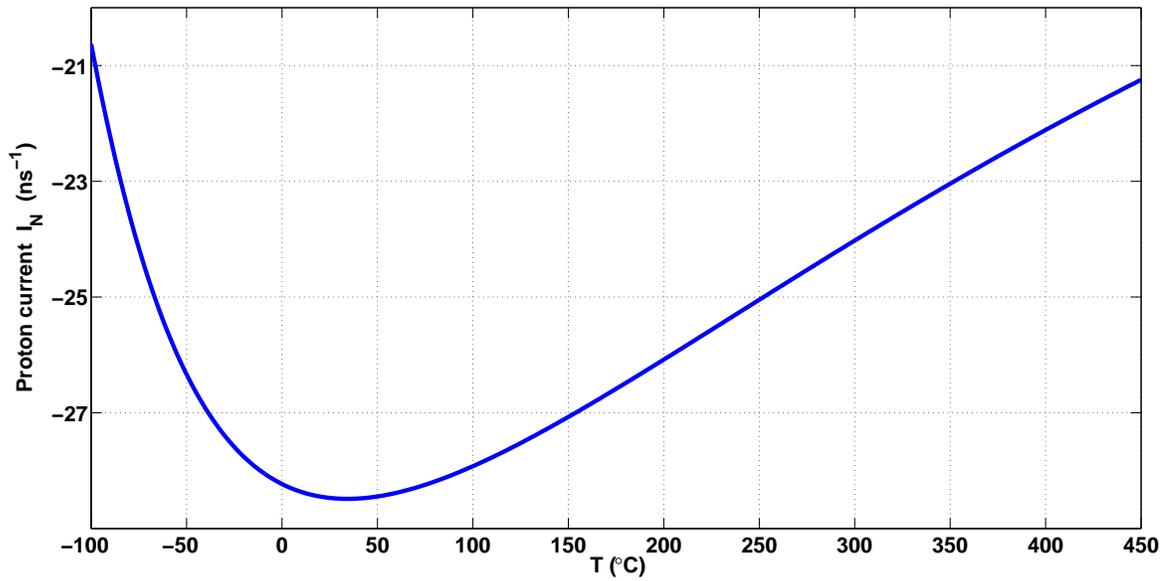}
\vspace*{0cm} \caption{ Proton current $I_N$ as a function of temperature $T$ for $E_2=850$ meV, $V_e=700 $ meV, $V_p=250$ meV.
The maximum value of the uphill proton current $|I_N|$ (which appears as a minimum in the plot) corresponds to the temperature $T
= 36.6^{\circ}$C.}
\end{figure}

\end{document}